\documentclass[pra,aps,showpacs,twocolumn,typeset,superscriptaddress]{revtex4}

\usepackage{amsmath}
\usepackage{amssymb}
\usepackage{graphicx}
\usepackage{epsfig}
\usepackage[latin1]{inputenc}
\begin{document}

\title{Quantum process reconstruction based on mutually unbiased basis}
\author{A. Fern\'andez-P\'erez}
\email{arturofe@ubiobio.cl}
\affiliation{Center for Optics and Photonics, Universidad de Concepci\'on, Casilla 4016, Concepci\'on, Chile}
\affiliation{Departamento de F\'isica, Facultad de Ciencias F\'isicas y Matem\'aticas, Universidad de Concepci\'on, Casilla 160-C, Concepci\'on, Chile}
\author{A. B. Klimov}
\affiliation{Center for Optics and Photonics, Universidad de Concepci\'on, Casilla 4016, Concepci\'on, Chile}
\affiliation{Departamento de F\'isica, Universidad de Guadalajara, Revoluci\'on 1500,
44410, Guadalajara, Jalisco, M\'exico}
\author{C. Saavedra}
\affiliation{Center for Optics and Photonics, Universidad de Concepci\'on, Casilla 4016, Concepci\'on, Chile}
\affiliation{Departamento de F\'isica, Facultad de Ciencias F\'isicas y Matem\'aticas, Universidad de Concepci\'on, Casilla 160-C, Concepci\'on, Chile}
\date{\today}

\begin{abstract}
We study a quantum process reconstruction based on the use of mutually unbiased projectors (MUB-projectors) as input states for a $D$-dimensional quantum system, with $D$ being a power of a prime number. This approach connects the results of quantum-state tomography using mutually unbiased bases (MUB) with the coefficients of a quantum process, expanded in terms of MUB-projectors. We also study the performance of the reconstruction scheme against random errors when measuring probabilities at the MUB-projectors.
\end{abstract}

\maketitle

\section{Introduction}
\label{introduction}

In physical implementations of quantum information processing, the ability for reconstructing the quantum process associated with a particular experiment is one of the most important tasks, which is mainly related to the determination of the average fidelity of a quantum process in the Hilbert space of states. The full characterization of the imperfection operations on a sequence of unitary transformations in quantum devices, and the effects of decoherence on the desired quantum evolution allow one to determine the fidelity of the quantum process. One of the more commonly used procedures is the so-called standard quantum process tomography (SQPT) \cite{Nielsenbook,Nielsen1997,Poyatos1997}, where the dynamics of a \textit{quantum black box}, described by a completely positive map $\mathcal{E}(\rho)$ \cite{Krausbook,Sudarshan1961}, is reconstructed. In this method, for a $D$-dimensional quantum system, one is required to prepare an ensemble of $D^{2}$ linearly independent input states $\left\{\rho _{k}=|\psi_{k}\rangle \langle \psi _{k}|\right\} _{k=0}^{D^{2}-1}$, subjecting each of them to the quantum process to be characterized, followed by standard quantum-state tomography on the output states, $\mathcal{E}(\rho_{k})$. The linearity of $\mathcal{E}$ relates the dynamical map to the experimental outputs via a linear system of equations which, after an inversion of a linear system of equations, allows to reconstruct the quantum-dynamical map.

Several experiments have been implemented by considering the above described reconstruction scheme, among them liquid-state NMR \cite{Nielsen1998}, light qubits \cite{James2001,OBrien2004}, atoms in optical lattices \cite{Steinberg2005}, cavity QED \cite{Brune2008}, trapped ions \cite{Riebe2006} and solid-state qubits \cite{Howard2006}. In all these experiments, the use of SQPT allows: the determination of the Kraus operators associated with the system's dynamics \cite{Howard2006}; estimation of the presence of decoherence mechanisms \cite{Kofman2009}; computing the experimental fidelity of a quantum gate \cite{Nielsen1998,OBrien2004,Riebe2006} and, in this way, we can have an estimation of how close the real dynamics are to the theoretical prediction. In SQPT only single-body interactions are required for the reconstruction procedure.

Recently, an alternative scheme for quantum process tomography (QPT) which uses an ancillary system has been implemented, which is the so-called assisted process tomography (AAPT) \cite{Ariano2001,Altepeter2003,Ariano2003}. In this scheme, an isomorphism between a quantum process and a quantum state in an extended Hilbert system is established \cite{Mohseni2008}. An input state, defined on the joint Hilbert space, passes through the map $\mathcal{E}$ and a projective measurement on the the output state is performed, thus reconstructing the quantum-dynamical map. These measurements can be implemented by using factorized measurement or joint measurements using mutually unbiased bases (MUBs). If a quantum map acts on a set of qubits, one ancilla is added to each qubit and joint measurements are performed on one system's qubit and on its ancilla. For this reason, for an $N$-qubits system $N$-ancillas are needed, and $5^{N}$ measurements must be carried out  for the reconstruction of the process \cite{Mohseni2008}. Another scheme that makes use of MUBs is the selective and efficient estimation of the parameters that characterize a quantum process, using MUBs for averaging the fidelities of appropriately modified channels \cite{Paz2008,Paz2009}. This scheme also needs an ancillary system for the determination of the off-diagonal terms of the process in its matrix representation.

In this article, we provide a new scheme for QPT, which is based on mutually unbiased basis quantum-state tomography (MUB-tomography) over the output states \cite{Klimov2008}. This procedure is valid for a Hilbert space with prime-power dimension, i.e., $D=p^{r}$ with $p$ and $r$ being a prime number and an integer, respectively \cite{Wootters1989}. For instance, it can be applied to a map acting on a single high dimensional system or on a multi-qubit system. In this scheme, two-body interactions are needed and do not require ancillary systems.

This paper has been organized as follows: In section \ref{MUB_QPT} we present the reconstruction of the QPT using MUBs, or simply MUB-QPT, showing the expansion of the quantum-dynamical map $\mathcal{E}$ in terms of MUB-projectors. Furthermore, we derive a relation between the expansion of the quantum map and the measurement of MUB-tomography, via a linear system of equations. In section \ref{numerical}, we present numerical simulations on the performance of the protocol against the presence of random errors in MUB-tomography. In the numerical simulation, we have considered quantum-dynamical maps describing a local decoherence channel and non-local quantum map. The article ends with a summary in Sec. \ref{summary}.

\section{QPT based on MUB-tomography}
\label{MUB_QPT}

For this purpose, let us consider the set of mutually unbiased projectors (MUB-projectors) for a $D$-dimensional system, whose dimension is a prime or a prime power number ($D=p^{r}$ with $p$ a prime number and $r$ a positive integer), where a set of mutually unbiased bases $|\psi _{m}^{(\gamma )}\rangle$ exist \cite{Wootters1989,Romero2005,Bandyopadhyay2002,Klimov2005,Ivanovic1981,Klimov2006}. For a $D$-dimensional system it has been found that the maximum number of MUBs cannot be greater than $D+1$ and this limit is reached if $D$ is prime \cite{Ivanovic1981} or power of prime \cite{Wootters1989}. Recently, it has been suggested that the existence of MUBs for other dimensions may well be related to the non-existence of finite projective planes of certain orders \cite{Saniga2004} or with the problem of mutually orthogonal Latin squares in combinatorics \cite{Wootters2006}. \\ The MUB-projectors are given by:

\begin{equation}
\mathcal{P}_{m}^{(\gamma )}=|\psi _{m}^{(\gamma )}\rangle \langle \psi
_{m}^{(\gamma )}|,\;m=1,...,D,\;\gamma =0,..,D,
\end{equation}
where $\gamma$ labels one of the $D+1$ families of MUBs and $m$ denotes one of the $D$ orthogonal states in this family. These projectors satisfy the following relation:
\begin{equation}
Tr\left( \mathcal{P}_{m}^{\left( \gamma \right) }\mathcal{P}_{n}^{\left( \beta \right)
}\right) =\delta _{\beta \gamma }\delta _{mn}+\frac{1}{D}\left( 1-\delta
_{\beta \gamma }\right).  \label{traza}
\end{equation}
Besides, these operators define a complete set of projection measurements, i.e., $\sum_{m=1}^{D}\mathcal{P}_{m}^{(\gamma)}=\hat{1}$ ($\hat{1}$ denotes the identity). The measured set of probabilities $p_{\gamma m}=Tr(\mathcal{P}_{m}^{(\gamma )}\rho )$ completely determines the density operator of the system, so that
\begin{equation}
\rho =\sum_{m=1}^{D}\sum_{\gamma =0}^{D}p_{\gamma m}\mathcal{P}_{m}^{\left(
\gamma \right) }-\hat{1}.  \label{rho}
\end{equation}
In a previous work \cite{Klimov2008}, we have shown an optimal protocol to obtain the probabilities $p_{\gamma m}$, for the case of having a multi-qubit system. The reconstruction scheme is based on minimizing the number of conditional logic gates (two-particle quantum operations) used in quantum-state tomography, due to the fact that all the MUBs contain non-factorizable bases. In this work, our main goal is to perform the quantum process reconstruction in a fully symmetrical form using MUB for input states, output state reconstruction and to expand the quantum dynamical map. The main motivation for doing this is the recent advancement in experimental implementation of MUBs quantum state tomography. It has been experimentally demonstrated quantum state tomography of two-qubit polarization using MUBs \cite{Adamson2010}. They demonstrate an improved state estimation when comparing with standard reconstruction. Besides, an experimental implementation of quantum tomography of 7- and 8-dimensional quantum systems has been reported \cite{Lima2011}, where higher dimensional quantum systems are encoded using the propagation modes of single photons, and MUBs projections have been implemented using programmable spatial light modulators.

To approach this problem, we begin by considering the general evolution of a $D$-dimensional quantum system described by a completely-positive linear map $\mathcal{E}\left( \rho \right) $, expressed in the so-called \textit{operator-sum representation},
\begin{equation}
\mathcal{E}(\rho )=\sum_{i}A_{i}\rho A_{i}^{\dag }  \label{Kraus}
\end{equation}
where $A_{i}$ are the Kraus operators of the system and satisfy $\sum_{i}A_{i}^{\dag }A_{i}\leq \hat{1}$ \cite{Nielsenbook}.

To expand the Kraus operators of the system, we consider the overcomplete basis of $(D^{2}+D)$ MUB-projectors, $\mathcal{P}_{m}^{(\alpha )}$ to expand the operators $A_{i}$,
\begin{equation}
A_{i}=\sum_{\alpha=0}^{D}\sum_{m=1}^{D} a_{im}^{(\alpha)} \mathcal{P}_{m}^{(\alpha )}.
\end{equation}%
Hence, the completely-positive linear map $\mathcal{E}\left( \rho \right)$ can be expressed in the following manner:
\begin{equation}
\mathcal{E}\left( \rho \right) =\sum_{\alpha , \beta=0}^{D}\sum_{m,n=1}^{D} \chi _{mn}^{\left(
\alpha ,\beta \right) }\mathcal{P}_{m}^{\left( \alpha \right) }\rho \mathcal{%
P}_{n}^{\left( \beta \right) }
\label{map}
\end{equation}
where $\chi _{mn}^{\left( \alpha ,\beta \right) }\equiv \sum_{i}a_{im}^{\left(
\alpha \right) }a_{in}^{\left( \beta \right) \ast }$ is the \textit{process matrix}. If we determine the $(D^{2}+D)^{2}$ quantities $\chi _{mn}^{\left( \alpha ,\beta \right) }$, we can reconstruct the quantum process in terms of the set of MUB-projectors $\mathcal{P}_{m}^{\left( \alpha \right) }$. Unless, the determination of the unital condition \cite{Perez2006}, $\sum_{i}A_{i}A_{i}^{\dag } \leq \hat{1}$ and the features of MUB-tomography \cite{Klimov2008}, mainly related with the expansion of the density operator in terms of MUB-projectors, as is shown in Eq (\ref{rho}), implies that we need to use the overcomplete basis of inputs to expand the Kraus operators and the density operator of the \textit{output states}. In this way we can present this scheme as an extension of the results obtained previously for quantum-state tomography using MUB. On the other hand, due to the complete set condition for each family of MUB (labelled by $\alpha$), the action of the quantum map over any input state $\mathcal{P}_{l}^{\left( \gamma \right) }$, according to Eq. (\ref{Kraus}) and the linearity of the super-operator, is:
\begin{eqnarray}
\mathcal{E}\left( \mathcal{P}_{l}^{\left( \gamma \right) }\right)
=\sum_{i}A_{i}A_{i}^{\dag }-\sum_{m\neq l}\mathcal{E}\left( \mathcal{P}%
_{m}^{\left( \gamma \right) }\right).
\label{linear}
\end{eqnarray}
Then, the action of the map on the remaining element $\mathcal{P}_{l}^{\left( \gamma \right) }$ ($m \ne l$) could be calculated by using the results of the super-operator $\mathcal{E}$ acting over the $D-1$ elements, $\sum_{m\neq l}\mathcal{E}\left( \mathcal{P}_{m}^{\left( \gamma \right) }\right)$. But the quantum dynamical map is unknown, and we can't predict the value of $\sum_{i}A_{i}A_{i}^{\dag}$. In other words, we don't know if the map is unital or not \cite{Perez2006}. For this reason, we need to determine the output state $\mathcal{E}\left( \mathcal{P}_{l}^{\left( \gamma \right) }\right)$ by quantum-state tomography. To summarize, for the reconstruction of the quantum process we need to prepare $(D^{2}+D)$ input states $\mathcal{P}_{l}^{\left(\gamma \right)}$ that correspond to the set of MUB-projectors. These states are sent through the quantum black-box and, after that, we carry out quantum-state tomography on the output states $\mathcal{E}( \mathcal{P}_{l}^{\left( \gamma \right) })$, based on mutually unbiased measurements \cite{Klimov2008}. We emphasize that the set of $(D^{2}+D)$ MUB-projectors used to expand the Kraus operators is the \textit{same set} used as input states for the quantum process.

We can expand any density operator using the overcomplete set of MUBs \cite{Klimov2008}. For this reason, the dynamical map associated with each input state, $\mathcal{E}( \mathcal{P}_{l}^{\left( \gamma \right) })$ can be expressed in the following way according to Eq. (\ref{rho}):
\begin{equation}
\mathcal{E}( \mathcal{P}_{l}^{\left( \gamma \right) })=\sum_{\epsilon =0}^{D} \sum_{q=1}^{D}p_{\epsilon q}^{\gamma,l}\mathcal{P}_{q}^{\left(
\epsilon \right) }-\hat{1},
\label{projectors}
\end{equation}%
where $p_{\epsilon q}^{\gamma,l}=Tr(\mathcal{P}_{q}^{(\epsilon )}\mathcal{E}( \mathcal{P}_{l}^{\left( \gamma \right) }))$ are the probabilities associated with each $\mathcal{P}_{q}^{(\epsilon )}$ projector after the dynamical map acts on the input state. These probabilities must be experimentally determined by using MUB-tomography.

Also, we can write the same dynamical map using Eq. (\ref{map}),
\begin{equation}
\mathcal{E}( \mathcal{P}_{l}^{\left( \gamma \right) })=\sum_{\alpha ,\beta
=0}^{D}\sum_{m,n=1}^{D}\chi _{mn}^{\left(
\alpha ,\beta \right) }\mathcal{P}_{m}^{\left( \alpha \right) }\mathcal{P}_{l}^{\left( \gamma \right) } \mathcal{%
P}_{n}^{\left( \beta \right) }.  \label{rhoj}
\end{equation}
Combining equations (\ref{projectors}) and (\ref{rhoj}) we
obtain
\begin{equation}
\sum_{\epsilon =0}^{D}\sum_{q=1}^{D}p_{\epsilon q}^{\gamma,l}\mathcal{P}_{q}^{\left(
\epsilon \right) }-\hat{1}=\sum_{\alpha ,\beta
=0}^{D}\sum_{m,n=1}^{D}\chi _{mn}^{\left(
\alpha ,\beta \right) }\mathcal{P}_{m}^{\left( \alpha \right) }\mathcal{P}_{l}^{\left( \gamma \right) } \mathcal{%
P}_{n}^{\left( \beta \right) }.
\end{equation}%
We make use of property (\ref{traza}), i.e., we apply the $\mathcal{P} _{s}^{\left( \eta \right) }$ projector on the above expression and take the trace on the whole expression obtaining
\begin{equation}
p_{\eta s}^{\left( \gamma ,l\right) }=\sum_{\alpha ,\beta
=0}^{D}\sum_{m,n=1}^{D} \chi
_{mn}^{\left( \alpha ,\beta \right) }Tr\left( \mathcal{P}_{m}^{\left( \alpha
\right) }\mathcal{P}_{l}^{\left( \gamma \right) }\mathcal{P}_{n}^{\left(
\beta \right) }\mathcal{P}_{s}^{\left( \eta \right) }\right).
\label{linsystem}
\end{equation}
We can observe from the equation (\ref{projectors}) that for each input state $\mathcal{P}_{l}^{\left( \gamma \right) }$ we can obtain a set of $D(D+1)$ probabilities from the state tomography. Then, with the experimental results of all the input states, we get a set of $(D^{2}+D)^{2}$ probabilities. Using this set,  $p_{\eta s}^{\left( \gamma ,l\right) }$, we write a linear system of equations (\ref{linsystem}) with all the terms of the $\chi_{mn}^{\left( \alpha ,\beta \right) }$ process matrix and the coefficients associated with the corresponding probability of the set, given by $Tr\left( \mathcal{P}_{m}^{\left( \alpha\right) }\mathcal{P}_{l}^{\left( \gamma \right) }\mathcal{P}_{n}^{\left(\beta \right) }\mathcal{P}_{s}^{\left( \eta \right) }\right)$. Clearly, as is shown in Eq. (\ref{linear}), not all the Eqs. that arise from (\ref{linsystem}) are linearly independent. This is not a problem, because the inversion of the linear system is guaranteed in terms of the generalized inverse, as we show in Sec. (\ref{numerical}), even when only $D^{4}$ Eqs. are linearly independent. Also, there is the possibility to use only $D^{2}$ MUB-projectors and the problem of QPT is solved, but in that case the extension of our previous result, shown in Ref. \cite{Klimov2008} related to the reconstruction of quantum process, is not clear and that is the reason to make a protocol based on an overcomplete basis set. 

Now, for simplicity we assume that the probabilities $p_{\eta s}^{\left( \gamma ,l\right) }$ and the $\chi_{mn}^{\left( \alpha ,\beta \right) }$ coefficients can be expressed as vectors ($\vec{p}$ and $\vec{\chi}$, respectively) and that the values given by $Tr\left( \mathcal{P}_{m}^{\left( \alpha \right) }\mathcal{P}_{l}^{\left( \gamma \right) }\mathcal{P}_{n}^{\left(\beta \right) }\mathcal{P}_{s}^{\left( \eta \right) }\right)$ can be arranged in a $(D^{2}+D)^{2}$-dimensional complex square matrix $\mathbf{\hat{\beta}}$.

Let us consider the inversion of equation (\ref{linsystem}). If we define $Tr\left( \mathcal{P}_{m}^{\left( \alpha
\right) }\mathcal{P}_{l}^{\left( \gamma \right) }\mathcal{P}_{n}^{\left(
\beta \right) }\mathcal{P}_{s}^{\left( \eta \right) }\right) \equiv \beta _{\alpha \beta ,\gamma \eta }^{mn,ls}$, it follows that, for each $p_{\eta s}^{\left( \gamma ,l\right)}$,
\begin{eqnarray}
p_{\eta s}^{\left( \gamma ,l\right) }=\sum_{\alpha ,\beta ,m,n}\chi
_{mn}^{\left( \alpha ,\beta \right) }\beta _{\alpha \beta ,\gamma \eta }^{mn,ls},
\label{linsystem2}
\end{eqnarray}
namely, $\vec{p} = \hat{\beta} \vec{\chi}$.To make a proper inversion of Eq. (\ref{linsystem2}), let us consider $\mathbf{\hat{\kappa}}$, the generalized inverse of matrix $\mathbf{\hat{\beta}}$ \cite{Israelbook}, which fulfills:
\begin{eqnarray}
\beta _{\alpha \beta ,\gamma \eta }^{mn,ls}=\sum_{\substack{ r,t,x,y \\ \mu
\,\nu ,\omega \,,\theta }}\beta _{\alpha \beta ,\gamma \eta }^{rt,xy}\kappa
_{rt,xy}^{\mu \nu ,\omega \theta }\beta _{\mu \nu ,\omega \theta }^{mn,ls}.
\label{pseudoinverse}
\end{eqnarray}
Hence, $\chi_{mn}^{\left( \alpha ,\beta \right) }$ is defined as,
\begin{eqnarray}
\chi _{mn}^{\left( \alpha ,\beta \right) }\equiv \sum_{\gamma ,\eta
,l,s}\kappa _{mn,ls}^{\alpha \beta ,\gamma \eta }p_{\eta s}^{\left( \gamma
,l\right) }.
\label{solution}
\end{eqnarray}
In this manner, the problem of QPT can be solved by finding the generalized inverse $\mathbf{\hat{\kappa}}$ of the complex matrix $\mathbf{\hat{\beta}}$, and finally solving  $\vec{\chi}=\mathbf{\hat{\kappa}}\vec{p}$.

In general, the determination of probabilities in tomographic reconstruction of output states is subject of experimental errors ($p_{\eta s}^{(\gamma,l)} \rightarrow \tilde{p}_{\eta s}^{(\gamma,l)}=p_{\eta s}^{(\gamma,l)}+\epsilon_{\eta s}^{(\gamma,l)}$, where $\epsilon_{\eta s}^{(\gamma,l)}$ denotes the error for each output state), which can give origin to unphysical process matrix $\chi_{mn}^{(\alpha,\beta)}$ when inverting the linear system, Eq. (\ref{linsystem2}). This problem can be overcome in a similar way than in the quantum-state tomography \cite{James2001} by minimizing a deviation function $f(\vec{t})$, where $\vec{t}$ is a general parametrization of the physical matrix process coefficients $\tilde{\chi} _{mn}^{\left( \alpha ,\beta \right) }(\vec{t})$ \cite{OBrien2004,Howard2006}. In our case, this function can be written as:

\begin{eqnarray}
f(\vec{t})&=&\left|\sum_{\alpha,\beta=0}^{D}\sum_{m,n}^{D}\tilde{\chi} _{mn}^{\left( \alpha ,\beta \right) }(\vec{t})-\chi_{mn}^{(\alpha,\beta)}\right|^{2} \\ &+& \sum_{\gamma,l} \lambda^{(\gamma)}_{l}\left|\sum_{\alpha,\beta=0}^{D}\sum_{m,n}^{D}\tilde{\chi} _{mn}^{\left( \alpha ,\beta \right) }(\vec{t})Tr\left({\cal P}_{n}^{(\beta)}{\cal P}_{m}^{(\alpha)}{\cal P}^{(\gamma)}_{l}\right)- 1 \right|^{2}. \notag
\end{eqnarray}
The second term in the right hand side is obtained by summing on index $s$ in Eq. (\ref{linsystem2}) and $\lambda^{(\gamma)}_l$ are the Lagrange multipliers corresponding to each input state. The parametrization $\vec{t}$ is chosen in a such a way of preserving both the hermiticity and the positivity of the map.

The advantage of mutually unbiased basis quantum-state tomography (MUB-Tomography) is related with the ability of both to maximize information extraction per measurement and to minimize redundancy. Hence, the estimation of the output state has several advantages with the use of MUB-Tomography. For instance, there is no need to invert a linear system of equations, as in standard quantum-state tomography \cite{Adamson2010,Lima2011}. Our main purpose is to extend this advantage to quantum-process reconstruction by having a fully symmetrical form using MUBs, namely, using MUBs for input states, for output state reconstruction and for expanding the quantum dynamical map.

In the next section, we give an explicit example of this protocol for local and non-local processes in $D=4$ and show how the performance of the protocol is affected when random errors are introduced into the obtained probabilities from quantum-state tomography, for local and non-local channels. 

\begin{center}
\begin{table}[tbp]
\caption{Five sets of three operators defining the MUB set for $D=4$} \label{operators}
\begin{tabular}{c||ccc}
$\alpha$ &  & Operators &  \\ \hline\hline
0 & $\hat{\sigma} _{z}\hat{1}$ & $\hat{1}\hat{\sigma} _{z}$ & $\hat{\sigma} _{z}\hat{\sigma} _{z}$ \\
1 & $\hat{\sigma} _{x}\hat{1}$ & $\hat{1}\hat{\sigma} _{x}$ & $\hat{\sigma} _{x}\hat{\sigma} _{x}$ \\
2 & $\hat{\sigma} _{y}\hat{1}$ & $\hat{1}\hat{\sigma} _{y}$ & $\hat{\sigma} _{y}\hat{\sigma} _{y}$ \\
3 & $\hat{\sigma} _{x}\hat{\sigma} _{y}$ & $\hat{\sigma} _{z}\hat{\sigma} _{x}$ & $\hat{\sigma} _{y}\hat{\sigma}
_{z}$ \\
4 & $\hat{\sigma} _{y}\hat{\sigma} _{x}$ & $\hat{\sigma} _{z}\hat{\sigma} _{y}$ & $\hat{\sigma} _{x}\hat{\sigma}
_{z}$
\end{tabular}
\end{table}
\end{center}

\section{Example in $D=4$}
\label{numerical}

Here we apply the above described scheme for the process reconstruction in the case of a four-dimensional Hilbert space, $D=4$, which could be associated to a two-qubit system. There are several approaches for the construction of mutually unbiased bases for this bipartite quantum system \cite{Wootters1989,Romero2005,Bandyopadhyay2002,Klimov2005,Ivanovic1981,Klimov2006}, which satisfies the prime-power condition for its construction, $2^{2}=4$. In this case, we will use the two-qubit operator set of table \ref{operators}, such that the common eigenstates constitute a MUB. Then, we can obtain $20$ MUB-projectors from Table \ref{operators} (four from each row).

The set of projectors $ \mathcal{P}_{l}^{\left( \gamma \right)}$ ($\gamma=0,..,4$ and $l=1,..,4$) will be the input states for testing the accuracy of this procedure for three fixed quantum processes; amplitude damping and depolarizing channel, as local decoherence mechanisms \cite{Nielsenbook}, and a controlled-NOT gate, $U_{CNOT}$. Our purpose is to show the effects on the process matrix \textbf{$\chi$} associated with each map, which arise from the presence of noise in the estimation of the probabilities $p^{(\gamma,l)}_{\eta s}$, in each quantum process considered.

The Kraus operators that represent local decoherence operations we denoted as $\hat{A_{ij}}$. They act individually at each qubit (labelled by $(i,j)$), so $\hat{A_{ij}}=\hat{A_{i}}\otimes \hat{A_{j}}$ and satisfy $\sum_{ij}\hat{A}_{ij}\hat{A}_{ij} ^{\dag}=\hat{1}$. These local processes, for a single qubit operation, have the following representation:
\begin{center}
\begin{tabular}{l||l}
Depolarizing Channel & Amplitude Damping \\ \hline
$A_{1}=\sqrt{1-\frac{3p}{4}}\hat{1}$ & $A_{1}=\frac{1}{2}\left( 1+\sqrt{%
1-\gamma }\right) (\hat{1}+\hat{\sigma}_{z})$ \\ 
$A_{2}=\sqrt{\frac{p}{4}}\hat{\sigma}_{x}$ & $A_{2}=\frac{\sqrt{\gamma }}{2}%
\left( \hat{\sigma}_{x}+i\hat{\sigma}_{y}\right) $ \\ 
$A_{3}=\sqrt{\frac{p}{4}}\hat{\sigma}_{y}$ &  \\ 
$A_{4}=\sqrt{\frac{p}{4}}\hat{\sigma}_{z}$ & 
\end{tabular}
\end{center}
The input states evolve in these local channels as $\mathcal{E}( \mathcal{P}_{l}^{\left( \gamma \right)})=\sum_{ij}A_{ij} \mathcal{P}_{l}^{\left( \gamma \right)} \left( A_{ij}\right) ^{\dag }$. We have chosen these channels because none of the input states $\mathcal{P}_{m}^{\left( \alpha \right) }$ are an eigenstate of the $A_{ij}$ operators and, for this reason, all the states will be affected by these decoherence mechanisms. This assumption will be not accomplished if, for example, we choose the bit-phase flip channel \cite{Nielsenbook}. In this case, the operators $A_{ij}$ associated with the channel are those in the third row of table \ref{operators}. This implies that the input states $\mathcal{P}_{m}^{\left( 2 \right) }$ will not be affected by this local decoherence mechanism, because they are eigenstates of all the operators that belong to the set $\alpha=2$ and, in this way, the set of three projectors $\mathcal{P}_{m}^{\left( 2 \right) }$ will be a decoherence-free subspace (DFS) for the bit-phase flip channel. We avoid this by choosing the amplitude damping and the depolarizing channel, because none of the 20 MUB-projectors obtained from the table \ref{operators} constitute a DFS.

On the other hand, the considered non-local operation is the controlled-NOT gate, that can be written as a coherent sum:
\begin{eqnarray}
U_{CNOT}=\frac{1}{2}\left( \hat{1}\otimes \hat{1}+\hat{1}\otimes \hat{\sigma}_{x}+
\hat{\sigma}_{z}\otimes \hat{1}-\hat{\sigma}_{z}\otimes \hat{\sigma}_{x}\right).
\end{eqnarray}
This process evolves the input states according to $\mathcal{E}( \mathcal{P}_{l}^{\left( \gamma \right) })=\left( U_{CNOT}\right) 
 \mathcal{P}_{l}^{\left( \gamma \right) }\left( U_{CNOT}\right) ^{\dag }$. We apply this set of operations to the whole set of input states $ \mathcal{P}_{l}^{\left( \gamma \right) }$, as is shown in  Eq. (\ref{Kraus}). Then, with the output states obtained, $ \mathcal{E}(\mathcal{P}_{l}^{\left( \gamma \right) })$, we calculate the corresponding probabilities associated with this dynamical map, $p_{\eta s}^{\left( \gamma , l \right) }=Tr(\mathcal{E(P}_{l}^{\left( \gamma \right) })\mathcal{P}_{s}^{\left( \eta \right) })$.

To have an estimation of the quality of this procedure, we have considered an error model for the probabilities $p_{\eta s}^{\left( \gamma ,l\right) }$, because in the experiments we always have to deal with small errors associated with these values. Then, our noisy probability $\tilde{p}_{\eta s}^{\left( \gamma , l\right) }$ is obtained by adding to the calculated probability a random value  $\zeta$ (uniformly distributed between 0 and 1) times a constant error parameter $\mu$:
\begin{equation}
\tilde{p}_{\eta s}^{\left( \gamma , l\right) }= p_{\eta s}^{\left( \gamma , l\right) }+\mu \zeta _{\eta s}^{\left( \gamma , l\right) },
\label{error}
\end{equation}
where $\epsilon_{\eta s}^{(\gamma,l)}=\mu \zeta _{\eta s}^{\left( \gamma , l\right) }$. This allows us to simulate a noisy measurement of the output states $\mathcal{E(P}_{l}^{\left( \gamma \right) })$, for the three different quantum operations considered. For each output state $\mathcal{E(P}_{l}^{\left( \gamma \right) })$ we normalize the new probabilities obtained according to Eq. (\ref{error}), for each measurement basis of the MUB-Tomography (labelled by $\eta$), that is, $\sum_{s}\tilde{p}_{\eta s}^{\left( \gamma , l\right) }=1$. With these values, we can calculate a noisy process matrix associated with the quantum channels considered, $\tilde{\chi} _{mn}^{\left( \alpha ,\beta \right) }$, which is obtained in an experiment with non-ideal measurements of the output states. We obtain this new process matrix by using Eq. (\ref{solution}) with the noisy probabilities on the right side,
\begin{eqnarray}
\tilde{\chi} _{mn}^{\left( \alpha ,\beta \right) }=\sum_{\gamma ,\eta,l,s} \kappa _{mn,ls}^{\alpha \beta ,\gamma \eta }\tilde{p}_{\eta s}^{\left( \gamma
,l\right) }.
\label{solution2}
\end{eqnarray}
For this purpose, we perform a numerical simulation with 100 steps for every value of $\mu$, each one representing the error model for the probabilities defined by Eq. (\ref{error}), i.e., it adds a set of random values into the probabilities of detection. The dimensionless error parameter $\mu$ takes values between $0.01$ and $0.15$, with increments of $0.01$. We will then obtain then a set of 100 noisy process matrices $\tilde{\chi} _{mn}^{\left(\alpha ,\beta \right) }$, for each value of $\mu$.\\
\begin{figure}[t]
\includegraphics[width=0.46\textwidth]{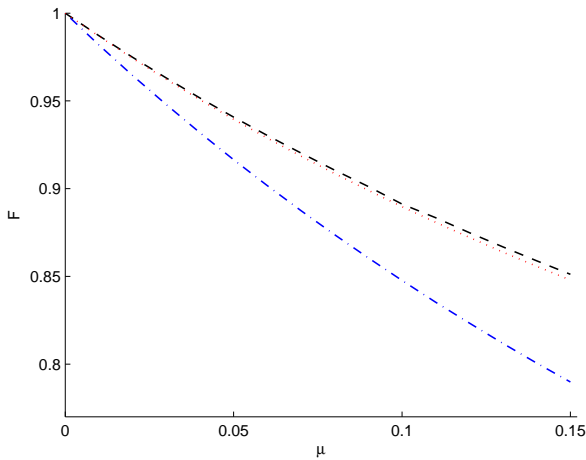}
\caption{\setlength{\baselineskip}{7pt}{\protect\footnotesize (Color online) Process fidelity $F$, for local and non-local quantum operations, when the process matrix $\chi$ is determined via noisy measurements in the MUB-Tomography, with error parameter $\mu$. Here, the fidelity in the determination of the process matrix associated with the non-local operation $U_{CNOT}$ is plotted with blue dashed-dot line. For the local channels, amplitude damping with $\gamma=0.4$ is plotted with red dotted line and the depolarizing channel with $p=0.1$ corresponds to the black dashed line.}}
\label{process}
\end{figure}

The process fidelity $F$ \cite{OBrien2004,Gilchrist2005} is a good way to estimate the quality in the determination of the dynamics with this basic error model, comparing the ideal process matrix $\chi_{mn}^{\left(\alpha ,\beta \right)}$ with each one of noisy matrices $\tilde{\chi} _{mn}^{\left(\alpha ,\beta \right) }$ obtained in the simulation. This quantity is defined as,
\begin{eqnarray}
F=\frac{Tr\left(\chi_{mn}^{\left(\alpha ,\beta \right)} \tilde{\chi} _{mn}^{\left(\alpha ,\beta \right) }\right) }{Tr\left( (\chi_{mn}^{\left(\alpha ,\beta \right)})^{2}\right) }
\end{eqnarray}
and take values between 0 and 1. Figure \ref{process} shows how the mean fidelity on the determination of the process matrix decreases when the error in the probabilities grows, as we expected. The mean fidelity is calculated using all the noisy matrices calculated $\tilde{\chi} _{mn}^{\left(\alpha ,\beta \right) }$ obtained in the simulation. Because of the introduction of the random errors, the precision in the estimation of the gate decays, but in the case of non-local gates, this behavior is more notorious. To explain this issue, we take one entangled input state and then apply the noisy non-local operation. Then, we calculate how the concurrence of the state changes when we introduce the random error. We observe that the concurrence decays in the same way as the process fidelity for that operation. Hence, the quantum correlations that the non-local operation induces are destroyed by the introduction of the random error, making the process fidelity $F$ decay more strongly than do the local ones.

The measurements in any set of MUBs contain non-factorizable bases, we require application of nonlocal gate operations. These operations currently can not be performed with unit fidelity. In our previous work \cite{Klimov2008}, we defined the \textit{physical complexity}, for each set of bases, as the number of CNOT gates needed for implementing projective measurements. Hence, for quantifying the amount of resources for implementing the QPT scheme, we use the physical complexity of the chosen set of MUBs.

In the case of MUB-Tomography, for a $D$-dimensional quantum system, the physical complexity for each set of MUB is denoted as $C_{\alpha}$ ($\alpha=0,...,D$). Then, for implementing projection measurements in $(D+1)$ bases, we need $C=\sum_{\alpha=0}^{D} C_{\alpha}$ CNOT Gates. Therefore, for each input state $\mathcal{P}_{l}^{\left( \gamma \right) }$ we need to perform $C$ non-local operations to carry out MUB-Tomography over the output state $\mathcal{E}\left( \mathcal{P}_{l}^{\left( \gamma \right) }\right)$. But, to prepare each input state (MUB-projectors) we also require a number of CNOT gates, given by the complexity of each MUB $C_{\alpha}$. As $D$ input states are prepared from each MUB; therefore the total number of CNOT gates required will be $(DC_{\alpha})$. Summing over the $(D+1)$ MUBs, then we need $(DC)$ non-local operations to prepare all the input states. Finally, we require a total number of $(DC^{2})$ CNOT gates for the whole QPT protocol.

Here, we assumed that the local operations are performed with unit fidelity. For a given dimension $D$ of the Hilbert space, only certain factorization structures are admitted \cite{Romero2005}, each one of them having a different complexity. Then, we need to minimize $C$ in order to carry out a lesser number of CNOT Gates. This is discussed in more detail in \cite{Klimov2008}.
 
\section{Summary}
\label{summary}

The choice of the set of MUB-projectors both as input states and as an operational basis, and the calculation of $p_{\eta s}^{(\gamma,l)}$ by the experimental data of MUB-Tomography, permit us to reconstruct an unknown quantum process, determining the process matrix $\chi_{mn}^{\left( \alpha ,\beta \right) }$, improving the way that standard QPT relates to the experimental outcomes with the parameters of the process matrix. We also can recover the Kraus operators for the system $A_{i}$, by the diagonalization of the process matrix. The simulations suggest good process fidelity, even when the values of the random errors $\mu \zeta _{\eta s}^{\left( \gamma ,l\right) }$ are similar to the theoretically calculated probabilities $p_{\eta s}^{\left( \gamma, l\right) }$, according to Eq. (\ref{error}), making this protocol robust against the presence of noise when determining probabilities. This protocol is valid in any Hilbert space with prime power dimension and where two body interactions are available, due to the fact that MUB-Tomography for all input states requires $(DC^{2})$ CNOT gates for its physical implementation, where $C$ is the physical complexity of the chosen MUBs. The case of not completely-positive maps \cite{Kuah2007}, the performance of this protocol compared with other procedures of QPT discussed in Sec. \ref{introduction}, in terms of the inaccuracy of estimated probabilities in the experiments, and the dimension of the considered Hilbert space, and how these features involve the estimation of the quantum process, are now left as open questions. A similar experimental setup used in Ref. \cite{Lima2011} can be used for a proof of principle implementation of this reconstruction scheme, where an extra spatial light modulator along the propagation path of single photons can be used for implementing different processes in a controlled way.

\textbf{Acknowledgments} We thank J. P. Paz for helpful discussions. This work was supported by grants Milenio ICM P06-67F and FONDECYT 1080383. A. Fern\'andez-P\'erez had support from CONICYT Scholarships and MECESUP FSM0605.

\end{document}